\documentclass[preprint]{rsl}
\usepackage{aasmacros}
\usepackage[sort&compress,square,numbers]{natbib}

\title{A new technique to measure noise parameters for global 21-cm experiments}
\author{Danny C. Price, Cheuk-Yu Edward Tong, Adrian T. Sutinjo, Nipanjana Patra, Lincoln J. Greenhill}

\begin{document}

\maketitle

%
%

\begin{abstract}
Radiometer experiments to detect 21-cm Hydrogen line emission from the Cosmic Dawn and Epoch of Reionization rely upon precise absolute calibration. During calibration, noise generated by amplifiers within the radiometer receiver must be accounted for; however, it is difficult to measure as the noise power varies with source impedance. In this letter, we introduce a convenient method to measure the noise parameters of a receiver system, which is practical for low-frequency receivers used in global 21-cm experiments.

\end{abstract}

\section{Introduction}

Numerous experiments (e.g. \cite{Bowman:2018,Price:2018,Singh:2022,deLeraAcedo:2022,Patra:2023}) seek to detect the global 21-cm signal from the Cosmic Dawn with radiometers across 30--250 MHz. Such a detection requires an exquisitely calibrated radiometer and a well characterized antenna. Any spectral features introduced by the radiometer may obfuscate, or be mistaken for, the expected $\sim$100mK amplitude absorption feature.

The EDGES experiment reported the presence of an apparent $\sim$500mK absorption feature in their calibrated data \cite{Bowman:2018}; however, there are concerns that this is partly or fully due to unmodelled systematics that have introduced a spectral feature \cite{Hills:2018,Singh:2019,Sims:2020}. In tension with the EDGES result, the SARAS-3 experiment has recently reported a significant non-detection of the absorption feature in their calibrated spectra \cite{Singh:2022}. Gaining a better understanding and characterization of systematics within global 21-cm experiments will be critical for breaking this tension.

One potential source of unwanted spectral features deserving more attention is the self-noise introduced by amplifiers within the radiometer. As the noise performance of a radiometer depends upon the impedance of the antenna it is connected to, a radiometer’s noise figure will differ when deployed in the field as compared to laboratory measurements with a 50$\Omega$ noise source. This difference is a source of error that must be accounted for.

To date, most global experiments have used a method involving a long coaxial cable to determine the magnitude of the difference, following the approach employed in EDGES \cite{Rogers:2012}. This method is primarily based upon Hu \& Weinreb \cite{Hu:2004}, but using the ``noise wave'' formulation of Meys \cite{Meys:1978}. Here, we introduce a new technique to characterize noise performance for global 21-cm experiments, based on the related “noise parameter” formulation commonly used in the microwave engineering community \cite{Haus:1960}.

Our technique, based on the approach given \cite{Price:2022}, only requires a standard Vector Network Analyzer (VNA) and calibration kit, a short cable, and a calibrated noise source. We refer the reader to \cite{Price:2022} for discussion of the mathematical and theoretical background. 

\section{Noise parameters}

The noise performance of a device-under-test (DUT) is commonly characterized by noise parameters: 
four real-valued terms from which noise characteristics
can be derived for any input impedance. Following \cite{Price:2022}, the noise temperature $T$ of
a 2-port DUT connected to a source with reflection coefficient $\Gamma_s$ can be expressed as:
\begin{align}
T(\Gamma_{s}) & =T_{{\rm min}}+T_{0} N \frac{\left|\Gamma_{s}-\Gamma_{{\rm opt}}\right|^{2}}{(1-\left|\Gamma_{s}\right|^{2})\left|1+\Gamma_{{\rm opt}}\right|^{2}}\label{eq:noise-param}
\end{align}
where the noise parameters are:
\begin{itemize}
\item $T_{{\rm min}}$ is the minimum noise temperature.
\item $\Gamma_{{\rm opt}}$ is the optimum reflection coefficient. 
\item $N$ is the minimum noise ratio.
\end{itemize}
Note that $T_0=290\,$K, $Z_0=50\,\Omega$, and $\Gamma_s$ is complex valued. We may treat the magnitude $\gamma_{\rm{opt}}$ and phase $\theta_{\rm{opt}}$ of $\Gamma_s$ as two real-valued noise parameters.

The noise parameters of a DUT may be determined by making at least 4 measurements of $T(\Gamma_s)$ when different source impedances are connected. Following Lane's method \cite{Lane:1969},  after the singularity removal detailed in \cite{Sutinjo:2020}, the four (real-valued) noise parameters can be found by casting the problem as a matrix equation:
\begin{align}
A\text{{\bf x}}&={\bf t}'\\
{\bf x}&=[a,b,c,d]^{T}\\
\mathbf{t'}&=(1-\gamma_{s_{i}}^{2})\mathbf{t}
\end{align}
Vector
${\bf t'}$ is formed from receiver measurements (entries in ${\bf t}$), matrix $A$ is formed from $\Gamma_{s}$ measurements, and we wish to find  (4$\times$1) noise parameter vector ${\bf x}$. The rows of matrix $A$ depend upon the formulation used; here, we will follow the formulation detailed in \cite{Hu:2004,Sutinjo:2020}:
\begin{align}
A_{i}&=\left[1-\gamma_{i}^{2},1,\gamma_{i}cos\theta_{i},\gamma_{i}sin\theta_{i}\right].\label{eq:Agamma}
\end{align}
and noise parameters are related to ${\bf x}=[a,b,c,d]^{T}$ by:
\begin{align}
T_{{\rm min}}&=a+\frac{b+\Delta}{2}\\
N &=\frac{\Delta}{4T_{0}}\\
\gamma_{{\rm opt}}&=\sqrt{\frac{b-\Delta}{b+\Delta}}\\
\theta_{{\rm opt}}&=tan^{-1}\left(\frac{-d}{-c}\right),
\end{align}
where $\Delta=\sqrt{b^{2}-c^{2}-d^{2}}$. Note that the minus signs in Equation 9 ensure the correct quadrant is returned.

\subsection{Approaches to source impedance selection}

\paragraph{Long cable approach:} To date, the main approach used in 21-cm experiments is to connect a long length of coaxial cable to the receiver so that the phase of a reflected noise waves wrap rapidly with frequency \cite{Hu:2004, Rogers:2012}. For 21-cm experiments, ``long'' lengths of 3--25\,m have been used \cite{Rogers:2012,Roque:2021}. The phase of an open cable's reflection coefficient is approximately
\begin{equation}
    \theta(f) = -4 \pi l / v_c f,
\end{equation}
so as the cable length $l$ increases, the phase will wrap faster, corresponding to circles around the edge of the Smith chart. If wraps are sufficiently fast---that is, faster than frequency variations of noise parameters---then well-spaced loci on the Smith chart can be selected to determine the underlying noise parameters. With this approach, a moving window of points with a least-squares fit can be used in lieu of the matrix method \cite{Hu:2004}. 

For the long-cable approach, higher frequency resolution requires longer cables. If the cable is too short, rapidly varying spectral structure will be missed. A length matched pair of open and shorted cables can be used so that the spacing on the Smith chart is constant across frequency .

\paragraph{OSLC approach:}
As shown in \cite{Price:2022}, noise parameters can be determined using the ``OSLC'' approach: an open ($\Gamma_{\rm{op}}\approx 1$), short ($\Gamma_{\rm{sh}}\approx-1$) and load ($\Gamma_{\rm{ld}}\approx0$) from a VNA calibration kit, and a shorted or open 1/8-wavelength length of cable ($\Gamma_{\rm{cbl}}\approx \pm 1j$). The OSLC approach can be used over a frequency range 0.2--1.8\,$f_0$, where $f_0$ is the frequency for which the cable is 1/8 of a wavelength for the transmission mode.  The loci of the OSLC points on a Smith chart form a ``well spread impedance pattern'', which has been shown to minimize measurement uncertainties \cite{Himmelfarb:2016}.

For 21-cm experiments, a 1/8-wavelength cable at a frequency $f_0=100\,$MHz (i.e. $\lambda_0=3\,$m) would be suitable to cover 20--180\,MHz. The cable would have a physical length $l=\lambda_0 v_c/8$, where $v_c$ is the velocity factor of the cable; for common dielectrics, $0.6<v_c<0.9$, so the required cable length is between 22.50--33.75\,cm. 

\paragraph{Comparison of approaches:} The OSLC approach uses a far shorter cable, and the frequency resolution of derived noise parameters depends only upon the receiver's channel bandwidth. Another advantage is that VNA calibration kits include physical models from which $\Gamma_s$ can be derived with higher precision than VNA measurements (see \cite{Monsalve:2016, Price:2022}). In contrast, the long-cable approach requires fewer source impedances to be connected, and thus fewer measurements, but much longer cables are needed if high frequency resolution solutions are required.

\section{Applying the OSLC method \label{sec:method}}

Here, we present a procedure to measure the noise parameters of a radiometric receiver using the OSLC approach. This procedure is similar to \cite{Price:2022}, but has been modified for the case where the receiver itself is the DUT\footnote{The main difference in derivation is that \textit{transducer} power gain must be used if the receiver is the DUT instead of \textit{available} power gain in eqn 35 of \cite{Price:2022}, if the DUT is connected in cascade to the receiver; see Chapter 11 of \cite{pozar2011microwave}.}. A summary of terms and measurements used in this section is provided in Table 1.

The OSLC method requires a calibrated noise source to generate ``hot'' (noise diode on) and ``cold'' (noise diode off) temperature references ($T_{\rm{hot}}$ and  $T_{\rm{cold}}$). The reflection coefficient must be measured for both states, and should satisfy $\Gamma_{\rm{hot}}\approx\Gamma_{\rm{cold}}$. 

To extract noise parameters of a radiometer, we need to form the source reflection coefficient matrix $A$, and measurement vector $\mathbf{t}'$. To form $A$, VNA measurements (or physical models) of the four source impedances (open, short, load cable) are plugged into equation 5. The measurement vector $\mathbf{t}'$, requires that four power measurements are made with the radiometer: one for each source impedance.

The power spectral density (PSD) measured by the radiometer when connected to a source with reflection coefficient $\Gamma_{s}$ is given by 
\begin{equation}
P_{s}(f, \Gamma_s)=D_{{\rm rx}}k_{B}\Delta f{G}_{{\rm rx}}(f, \Gamma_s)\left[T_{s} (f) +T_{{\rm n}}(f, \Gamma_s) \right]\label{eq:power-basic}
\end{equation}
where $k_{B}$ is the Boltzmann constant and $\Delta f$ is the noise
equivalent bandwidth per channel. $T_{n}$ is the receiver noise temperature when connected to $\Gamma_{s}$. $T_{s}$ is the source noise temperature, equal to the ambient temperature for passive components. $G_{{rx}}(f, \Gamma_s)$
is the transducer gain of the receiver; the factor ${D}_{{\rm rx}}$
represents to digital gain factors within the receiver (assumed to be linear). 

To calibrate, we define a scale factor $\alpha$:
\begin{align}
\alpha & =\frac{T_{{\rm hot}}-T_{{\rm cold}}}{P_{{\rm hot}}-P_{{\rm cold}}} = \frac{1}{D_{{\rm rx}}k_{B}\Delta f{G}_{{\rm rx}}(f, \Gamma_s)},
\end{align}
and similarly we define a mismatch factor
\begin{align}
M(\Gamma_{s}) = \frac{G_{\rm{rx}}(f, \Gamma_s)}{G_{\rm{rx}}(f, \Gamma_{\rm{ns}})}  & =\left(1-\left|\Gamma_{{\rm ns}}\right|^{2}\right)\frac{\left|1-\Gamma_{s}\Gamma_{{\rm rx}}\right|^{2}}{\left|1-\Gamma_{{\rm ns}}\Gamma_{{\rm rx}}\right|^{2}}.\label{eq:mismatch-1}
\end{align}
By doing so, we may form $\mathbf{t}'$ by:
\begin{equation}
{\bf t}'_{i}=\left(\alpha P_{s_{i}}M_{s_{i}}-\left(1-\left|\Gamma_{s_{i}}\right|^{2}\right)T_{s_{i}}\right),
\end{equation}
The noise parameter vector may then be recovered via
\begin{equation}
{\bf x} = A^{-1} {\bf t},
\end{equation}
after which noise parameters $T_{\rm{min}}$, $R_N$, $\gamma_{\rm{opt}}$ and $\theta_{\rm{opt}}$ are recovered via equations 6--9.

\begin{table}
\centering
\small
\caption{Summary of reflection coefficients and power spectral density (PSD) measurements required for the OSLC method. \label{tab:params}}

\begin{tabular}{cl}
  &  \\
\hline
\hline
$\Gamma_{\rm{rx}}$   & Refl. coefficient of radiometer receiver \\     
$\Gamma_{\rm{hot}}$  & Refl. coefficient of cal noise source (on).  \\
$\Gamma_{\rm{cold}}$ & Refl. coefficient of cal noise source (off).  \\
$\Gamma_{\rm{ns}}$   & Computed via $\Gamma_{\rm{ns}}= (\Gamma_{\rm{on}} + \Gamma_{\rm{off}}) / 2$\\
$\Gamma_{\rm{op}}$   & Refl. coefficient of open standard \\                                     
$\Gamma_{\rm{sh}}$   & Refl. coefficient of short standard \\     
$\Gamma_{\rm{ld}}$   & Refl. coefficient of broadband load standard \\     
$\Gamma_{\rm{cbl}}$  & Refl. coefficient of 1/8-wavelength cable \\   
\hline 
$P_{\rm{hot}}$  & Recv. PSD when cal noise source (on) connected  \\
$P_{\rm{cold}}$ & Recv. PSD when cal noise source (off) connected   \\
$P_{\rm{op}}$   & Recv. PSD when open standard is connected \\                
$P_{\rm{sh}}$   & Recv. PSD when short standard is connected \\   
$P_{\rm{ld}}$   & Recv. PSD when load standard is connected \\      
$P_{\rm{cbl}}$  & Recv. PSD when $\lambda$/8 cable is connected \\  
\hline
$T_{\rm{cold}}$ & Ambient temperature (for `cold' noise source (off)) \\
$T_{\rm{hot}}$ & Noise source effective temperature (i.e. ENR)  \\
\hline
\end{tabular}
\end{table}

\section{Application to HYPEREION}

We used the OSLC approach to measure the noise parameters of the prototype receiver for the HYPEREION system \cite{Patra:2023}. The HYPEREION system implements a two-channel, cross-correlation spectrometer; for simplicity, we only consider a single autocorrelation channel here. All required VNA and power spectra measurements (Table 1), were taken in a laboratory setting.

HYPEREION consists of a ``frontend module'', which performs initial signal conditioning, connected to the ``backend module'' and digital signal processor by 100\,m of coaxial cable. When deployed, the frontend module will be connected to the antenna and located in the field, and the backend module and digital system will be located in an electromagnetically shielded room.

We connected an open, short, load (from an Agilent 85052D calibration kit) and shorted coaxial cable to the HYPEREION frontend module, and recorded power spectra in each state, across 30--120\,MHz. A Keysight HP346B calibrated noise source was used to provide hot and cold reference states, and a Fieldfox N9915A VNA was used to measure reflection coefficients. Power spectra were generated using the HYPEREION digtal receiver, which is based on a 14-bit Signatek PX1500-2 digitizer.

Extracted noise parameters are shown in Figure 1.

\begin{figure}
\begin{centering}
\includegraphics[width=1.0\columnwidth]{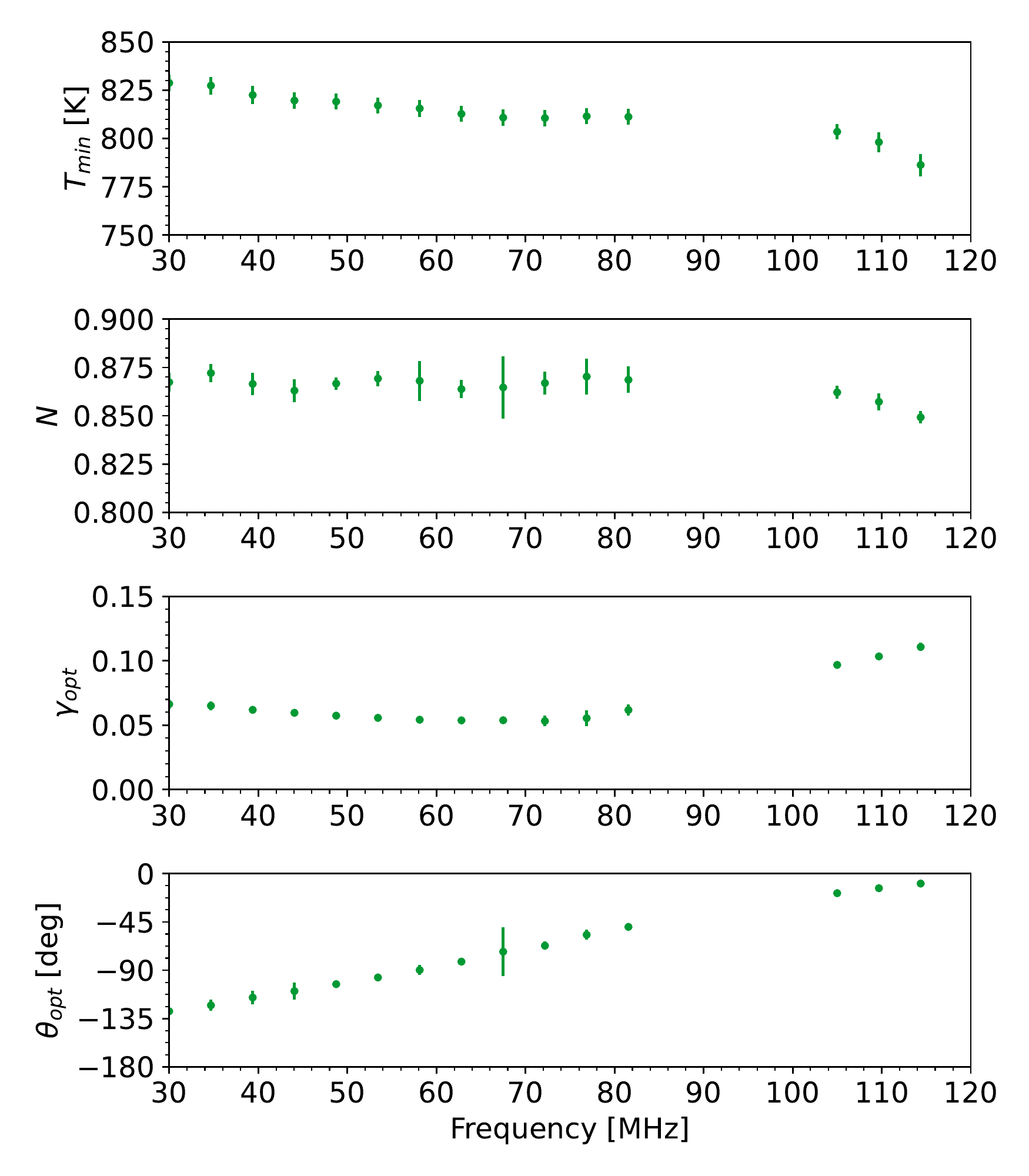}
\par\end{centering}
\caption{ Measured noise parameters for the HYPEREION prototype system using the OSLC method. Points between 88--108\,MHz have been flagged due to strong FM-band radio interference present in the data.}
\end{figure}

\section{Discussion and conclusions}

Here, we have introduced a noise parameter measurement approach that can be applied to  radiometer receivers used in 21-cm experiments. The approach presented here is a modified version of the approach detailed in \cite{Price:2022}, which provides extended details and discussion of noise parameter measurement techniques.

Many global 21-cm experiments switch between the antenna and a set of calibration references, and have existing methods to convert measured data into temperature (K). These internal references may be used in lieu of external open/short/load, as long as their reflection coefficients can be accurately modelled. Similarly, an experiment's existing calibration routines may be used in lieu of the procedure outlined in Section 3. 

The OSLC method is suitable for in-situ application (i.e. in the field) using a portable VNA. We suggest that future global 21-cm experiments should consider integrating OSLC source impedances within the radiometer.

\bibliography{IEEEabrv,references}
\bibliographystyle{IEEEtran}

\end{document}